\documentclass[a4paper,12pt]{article}

\usepackage{citesort}
\usepackage{amsmath}

\def \gsim
{\raisebox{-3pt}{$\>\stackrel{>}{\scriptstyle\sim}\>$}}

\def\beq{\begin{equation}}
\def\be{\begin{equation}}
\def\eeq{\end{equation}}
\def\ee{\end{equation}}
\def\bea{\begin{eqnarray}}
\def\eea{\end{eqnarray}}

\def\bq{\begin{quote}}
\def\eq{\end{quote}}

\def\gappeq{\mathrel{\rlap {\raise.5ex\hbox{$>$}}
{\lower.5ex\hbox{$\sim$}}}}
\def\lappeq{\mathrel{\rlap{\raise.5ex\hbox{$<$}}
{\lower.5ex\hbox{$\sim$}}}}
\def\Toprel#1\over#2{\mathrel{\mathop{#2}\limits^{#1}}}

\allowdisplaybreaks

\begin{document}

\begin{titlepage}
\vspace*{-2cm}
\begin{flushright}
ROME1/1440-06
\end{flushright}

{\Large
\begin{center}
{\bf Resummed Mass Distribution for Jets Initiated by Massive 
Quarks~$^*$}
\end{center}
}
\vspace{.5cm}

\begin{center}
{U.~Aglietti~$^{a}$\,,~~~~L.~Di~Giustino~$^{b}$\,,~~~~G.~Ferrera~$^{a}$\,,
~~~~L.~Trentadue~$^{b}$}
\\[7mm]
{$^{a}$\textit{Dipartimento di Fisica, Universit\`a di Roma ``La Sapienza''
\\
and
\\
INFN Sezione di Roma, Roma, Italy }}
\\[4mm]
{$^{b}$\textit{Dipartimento di Fisica,~ Universit\`a di Parma,
\\
and
\\
INFN,~ Gruppo Collegato di Parma,~ Parma,~ Italy}}\\
[10pt] \vspace{1.5cm}
\begin{abstract}
We resum the invariant mass distribution of jets initiated by massive quarks
in next-to-leading logarithmic approximation and beyond in heuristic way. 
We find that the inclusion of mass terms, in the $N$-moment space, 
results in the universal factor
\be
\!\!
\delta_N(Q^2; m^2)=
\mbox{\large \textit e}^{\, {\displaystyle \int_0^1} \!\!
{\displaystyle d z } {\displaystyle  { \frac{ z^{ \frac{m^2}{Q^2} (\!N-1\!) }
\!-\!1
}{ 1 - z}} } \left\{ -\! { \displaystyle
\int_{m^2(1-z)^2}^{m^2 (1-z)}} { \displaystyle
\frac{dk_{\perp}^2}{k_{\perp}^2}}  {\displaystyle
A}{\left[\alpha_S\left(\! k_{\perp}^2\!\right)\!\right]} -{\displaystyle
B}{\left[\alpha_S\left( \!m^2 (1-z)\! \right)\!\right]} +{\displaystyle
D}{ \left[\alpha_S\left( \!m^2 (1-z)^2\! \right)\!\right]} \!\right\}
}\!,\!
\nonumber
\ee
taking into account dead-cone effects and soft radiation characteristic
of massive charges. This factor multiplies the massless jet distribution function $J_N(Q^2)$.
In the above equation the variable $N$ is rescaled by the mass correction parameter $ m^2/Q^2 \ll 1$
with respect to the standard massless case, being $m$ the quark mass and $Q$ the hard scale.
The functions $A(\alpha_S)$ and $B(\alpha_S)$ 
appear with a minus sign suppressing collinear effects at very large $N \gsim Q^2/m^2$, as expected. 
In the same region, 
soft radiation not collinearly enhanced, characteristic of on-shell massive charges, makes its
appearance with the function $D(\alpha_S)$.
\noindent
Phenomenological applications, such as the resummation
of $b\to c l \nu $ decay spectra or the inclusion of beauty mass
effects in $t\to b W$ decays, are briefly sketched.

\end{abstract}
~\\
$^*$ {\large \itshape This work is dedicated to the memory of Jiro Kodaira.}
\end{center}
\end{titlepage}

\noindent

\section{Introduction}

The structure of hadronic final states at high energy is largely determined by
the infrared (soft and collinear) divergencies occurring in real 
contributions to QCD cross sections. The singularities cancel 
with those coming from virtual corrections in sufficiently inclusive observables, 
but leave often large residual logarithmic effects \cite{qcdgeneral}. 
Soft singularities mean  in physical terms a high probability of soft gluon emission:
\beq
dP \, \sim \, \alpha \, \frac{dE}{E},
\eeq
where $\alpha \, \equiv \, \alpha_S$ with $E$ the gluon energy and $Q$ the hard scale 
($\Lambda_{QCD} \ll E \ll Q$, $\Lambda_{QCD}$ being the hadronic scale).
Because of such singularities, one has to think to a quark or a gluon as a ``dressed'' parton,
which is always accompanied by a cloud of virtual soft gluons.
In a scattering process, part of the cloud ``detaches'' and converts itself
into real soft radiation, observed as low-energy hadrons.
Collinear singularities in physical terms mean high probability of collinear
configurations: 
\beq
\label{collconf}
dP \, \sim \, \alpha \, \frac{d\theta^2}{\theta^2},
\eeq
where $\theta$ is the emission angle.
Because of them, a massless quark or a gluon evolves into an
ensemble of collinear partons, which show up in the detectors as
a jet of hadrons. 
In the case of a {\it massless} parton, 
there is an overlap of soft and collinear regions producing the well-known
(leading) double-logarithmic effects:
\beq
dP \, \sim \, \alpha \, \frac{dE}{E} \, \frac{d\theta^2}{\theta^2}.
\eeq
Furthermore, there is {\it hard collinear} emission related to contributions
of the form (\ref{collconf}).
In the case of a {\it massive} parton --- which we assume to be observed in a
reference frame where it is fast moving --- two main changes occur in the QCD radiation.
The first is the well-known ``dead cone'' effect, according to which little radiation
is emitted in the forward direction:
\beq
dP_m \, \sim \, \alpha \, \frac{d\theta^2}{ \theta^2 \, + \, m^2/Q^2 }
\, \sim \,  \alpha \, H\left( \theta^2 - \frac{m^2}{Q^2} \right) \, \frac{d\theta^2}{ \theta^2 },
\eeq
where $H$ is the step function.
The second effect is related to the fact that a massive parton
radiates soft quanta {\it isotropically} in its {\it rest frame}:
\beq
\label{softmassive}
dP_m \, \sim \, \alpha \, \frac{dE}{E}.
\eeq
These quanta, being radiated in any space direction, cannot be ascribed in a natural way 
to any jet in the event. 
They are related to the ``classical'' chromo-electric field of a static color charge.
By means of a Lorentz transformation, we obtain a fast-moving parton, 
which is now accompanied by a boosted Coulomb field, i.e.
by soft radiation restricted to a forward cone of angular size 
$\theta \sim 1/ \gamma_L \sim m/Q$. 
As a consequence, when we consider the structure of a jet initiated by a parton with
a small mass, we expect {\it missing} collinear radiation compared to the massless case and 
{\it additional} soft radiation related to the boosted Coulomb field.
A non-vanishing quark mass, $m \ne 0$, produces, as we are going to show in detail, 
a specific sub-structure, which is expected to be universal on the basis of
physical intuition: small angle emissions only are involved, which
can be ascribed to a specific jet in the event.
In general, whenever logarithmically enhanced effects are encountered, 
one expects a factorized structure.  
As we are going to show in detail in this note, all these expectations turn
out to be correct.

In the massless case, it is convenient to introduce the
``jet function'' $J(y; Q^2)$, which gives the probability that a massless parton produced
in a hard process characterized by the scale $Q$ fragments into a hadronic
jet of mass squared \cite{catanitrentadue,Catani:1990rr,catcac}
\beq
\label{defy}
m_X^2 \, = \, y \, Q^2.
\eeq
In first order, its expression reads:
\beq
\label{eq7}
J(y; \, Q^2) \, = \, \delta(y) \, - \,A_1 \, \alpha \, \left( \frac{\log y}{y} \right)_+
\, + \, B_1\, \alpha \, \left( \frac{1}{ y } \right)
\, + \, O\left(\alpha^2\right),
\eeq
where $A_1$ and $B_1$ are coefficients whose explicit expressions will be given later
and the plus distributions are defined as:
\beq
P(y)_+ \, \equiv \,
\lim_{\epsilon \to 0^+}
\left[
\theta(y-\epsilon) P(y) \, - \,
\delta(y-\epsilon) \int_\epsilon^1 P(y') dy'
\right].
\eeq
Note that in the free limit, $\alpha \to 0$, the jet has a trivial structure and the
mass distribution reduces to a spike corresponding to the (zero) parton mass.
If $y \ll 1$, the coefficients of $\alpha$ in eq.~(\ref{eq7}) become large, making a truncated
perturbative expansion unreliable and asking for a resummation to all orders. 
The latter is systematically performed by going to $N-$space, i.e. by
considering 
\beq
J_N(Q^2) \, = \, \int_0^1 dy \, (1-y)^{N-1} \, J(y; \, Q^2).
\eeq
The resummed expression reads \cite{catanitrentadue,Catani:1990rr,catcac}:
\beq
J_N(Q^2) \, = \, \, \exp \,
\int_0^1 d y \, \frac{ (1-y)^{N-1} - 1 }{y}
\Bigg\{
\, \int_{ Q^2 y^2 }^{ Q^2 y } \frac{dk_{\perp}^2}{k_{\perp}^2} 
A\left[\alpha\left(k_{\perp}^2\right)\right]
\, + \, B \left[\alpha\left( Q^2 y \right)\right] \,
\Bigg\}.
\eeq
The functions $A(\alpha)$ and $B(\alpha)$, generalizing the first-order expressions
$A_1\alpha$ and $B_1\alpha$ respectively, describe small-angle {\it soft} and {\it hard} 
gluon emission off a massless parton respectively (see later).

In the massive case, one can define a {\it generalized} jet function
$J(y; \, Q^2, m^2)$, where $y$ is now defined as (cfr. eq.~(\ref{defy})):
\beq
\label{eqdefy}
y \, = \, \frac{m_X^2 - m^2}{Q^2 - m^2}.
\eeq
To first order, the massive jet function reads:
\bea
\label{splithard}
J(y; \, Q^2, m^2) &=& \delta(y) 
\, + \, \alpha
\Bigg[
\left(
- \,A_1 \, \log y
\, + \, B_1
\right) \, \frac{\theta(y - r)}{y}
\, +
\nonumber\\
&+& 
\left(
- A_1 \, \log r
\, + \, D_1 
\right) \, \frac{ \theta(r - y) }{ y }
\Bigg]_+
\, + \, O\left(\alpha^2\right),
\eea
where $\theta$ is the step function and $D_1$ is a coefficient 
which will be derived in the next section.
We have defined the mass correction parameter
\beq
\label{defr}
r \, \equiv \, \frac{m^2}{Q^2} \, \ll \, 1.
\eeq
Note that $r$ is quadratic in $m$ and that we always assume that the quark mass
is much smaller than the hard scale,
$m \, \ll \, Q$, is in order to have fast-moving charges and to preserve a jet
structure
\footnote{We are not interested to the case in which the quark mass
is close to the hard scale, $m \, \approx \, O(Q)$,
as this case corresponds to a non-relativistic motion of the color charges
which does not give rise to jets.}.

According to eq.~(\ref{splithard}), we can identify two different kinematical
regions, which can be defined with logarithmic accuracy:
\begin{enumerate}
\item
{\it high} jet mass: $y \, \gg \, r$ or, equivalently, $m_X - m \, \gg \, m$.
The quark mass $m$ can be neglected and the collinear
region produces $\log y$ terms;
\item
{\it low} jet mass: $y \, \ll \, r$ or, equivalently, $m_X - m \, \ll \, m$. 
The quark mass screens the collinear singularity and produces $\log r$ terms.
\end{enumerate}
Mass effects are more easily looked at by considering the 
partially integrated jet rate:
\beq
R(y; \, Q^2, m^2) \, \equiv \, \int_0^y J(y'; \, Q^2, m^2)  \, d y'.
\eeq
In the high-mass region:
\begin{equation}
R(y) \, = \, 1 \, - \, \int_y^1 J(y'; \, Q^2, m^2) \, d y'
\, = \, 1 \, - \, \frac{1}{2} A_1 \alpha \log^2 y \, + \, B_1 \, \alpha \, \log y
~~~~~~~~~~ (y \, > \, r),
\end{equation}
while in the low-mass region:
\begin{equation}
R(y) \, = \, 1 \, - \, A_1 \, \alpha \, \log y \, \log r
\, + \, \frac{A_1}{2} \, \alpha \, \log^2 r
\, + \, D_1 \alpha \log y \, + \, (B_1 - D_1) \alpha \log r
~~~~~~~~~~~~ (y \, < \, r).
\end{equation}
For $r \ll 1$ mass logarithms become large and need to be resummed
to all orders.
We will show that mass terms can be relegated into a factor which takes into
account the effects discussed above. The jet function can be
factorized in moment space as: 
\beq 
\label{univ} 
J_N(Q^2; \, m^2)
\, = \, J_N(Q^2) ~ \delta_N(Q^2; \, m^2), 
\eeq
where the mass-correction factor $\delta_N(Q^2; \, m^2)$ reads:
\bea
\label{maineq}
\hskip -1truecm
\delta_N(Q^2; m^2) &=& \exp
\int_0^1 d y \frac{ (1-y)^{ \, r \, (N-1)} - 1 }{y}
\Bigg\{
 - \int_{ m^2 y^2 }^{ m^2 y } \frac{dk_{\perp}^2}{k_{\perp}^2} A\left[\alpha\left(k_{\perp}^2\right)\right]
 - B \left[\alpha\left( m^2 y \right)\right]
+ \nonumber\\
&&~~~~~~~~~~~~~~~~~~~~~~~~~~~~~~~~~~~~~~~~~~~~~~~~~~~~~~~~~~~~~
\, + \, D \left[\alpha\left( m^2 y^2 \right)\right]
\Bigg\}.
\eea
The function $D(\alpha)$, generalizing the first-order term $D_1\alpha$, 
describes soft radiation not collinearly
enhanced off a nearly on-shell massive quark (see later).
We derive our main result, eq.~(\ref{maineq}), with some rigour
in next-to-leading logarithmic (NLL) approximation and we conjecture 
its general validity beyond NLL.
Eq.~(\ref{maineq}) has a simple physical
interpretation: the parameter $N-1$ is multiplied by $r$ on the
r.h.s., implying that mass effects become ``visible'' only for large
\beq
N \, \gsim \, \frac{Q^2}{m^2} \, \gg \, 1.
\eeq
In this case, there is enough resolution to ``see'' the quark mass, which tends to
suppress the collinear effects, related to the $A$ and $B$ terms.
At the same time, soft radiation not collinearly enhanced,
described by the function $D$ and characteristic of
massive partons, does appear.
Let us also note that, since the jet mass is an infrared (i.e. soft and collinear) safe quantity,
$\delta_N = 1$ for $r = 0$.

For clarity's sake, we derive the general formula (\ref{maineq}) in sec.~\ref{secbsgamma} 
for a specific physical process first, namely the radiative decay
\beq
\label{raddec}
B \to X_s + \gamma
\eeq
by keeping $m_s \, \ne \, 0$
\footnote{
Let us note that the inclusion of strange mass effects is a rather
academical problem because even with a large value of the constituent mass
$m_s \, = \, 0.5$ GeV, the correction parameter is very small: $r = (m_s/m_b)^2 \approx 10^{-2}$.
Furthermore, strange mass effects become visible for
$x_{\gamma} \, \equiv \, 2 E_{\gamma}/m_B \, = \, 1 - m_{Xs}^2/m_B^2 \, \gsim \, 0.99$,
where resonance effects due to $K$, $K^*$ peaks, etc. --- intrinsically
beyond perturbative QCD --- are substantial.
}.
In sec.~\ref{generalsec} we generalize the results of the previous section
to the case of a process involving one or more jets initiated by quarks with a small mass
compared to the hard scale.
Finally in sec.~\ref{conclusionsec} we present the conclusions of our study
and we discuss natural applications and developments.

\section{Mass Distribution in $B \to X_s + \gamma$ for $m_s \neq 0$}
\label{secbsgamma}

In this section we compute the invariant hadron mass distribution
in the decay (\ref{raddec}), having the fixed-order expansion:
\begin{equation}
\label{aggiunta}
\frac{1}{\Gamma} \frac{d\Gamma}{dy} \, = \,
\delta(y) 
\, + \, \frac{1}{\Gamma} \frac{d\Gamma^{(1)}}{dy} 
\, + \, \frac{1}{\Gamma} \frac{d\Gamma^{(2)}}{dy} 
\, + \, \cdots, 
\eeq 
where $1/\Gamma \, d\Gamma^{(n)}/dy \, = \, O(\alpha^n)$ 
and $\Gamma$ is the corrected inclusive width.
The massless case has been considered for example in ref.~\cite{sghedoni}. 

\subsection{Single-gluon distribution}
\label{onegluon}

In order to include the quark-mass corrections to 
eq.~(\ref{aggiunta}), let us start from the eikonal current
containing explicitly the mass terms.
According to standard factorization formulae, the soft terms in
the amplitude for 
\beq 
b \, \to \, s \, + \, g \, + \, \gamma 
\eeq
are factorized by the eikonal current 
\beq 
J^{\mu}(k) \, = \, i g \, \left( {\bf T_b} \, \frac{p_b^{\mu}}{p_b\cdot k} \, - \, 
{\bf T_s} \, \frac{p_s^{\mu}}{p_s\cdot k} \right), 
\eeq 
where $g$ is the coupling in the QCD lagrangian, $k^{\mu}$ is the soft gluon momentum
and ${\bf T_b}$ and ${\bf T_s}$ are the color generators for the
beauty and the strange quark respectively \cite{Catani:2002hc}.
Since the photon --- in general the probe --- carries no color, we
have color conservation along the quark line: \beq {\bf T_b \, =
\, T_s}. \eeq The square of the eikonal current reads: \footnote{
For simplicity's sake, we compute the square in Feynman (i.e.
covariant) gauge, but the result is gauge invariant because the
eikonal current is conserved: $k_{\mu} J^{\mu}(k) = 0$. } \bea
\label{square} - \, g_{\mu\nu} \, J^{\mu}(k) \, J^{\nu}(k)^* &=&
4\pi \alpha C_F \left[ \frac{2 \left( p_b\cdot p_s \right)
}{\left(p_b\cdot k\right) \, \left(p_s\cdot k\right)} \, - \,
\frac{m_b^2}{\left(p_b\cdot k\right)^2} \, - \,
\frac{m_s^2}{\left(p_s\cdot k\right)^2} \right]
\\
&=&
4 \pi \alpha \frac{C_F}{E_g^2}
\left\{
\frac{1+r}{1-r} \, \frac{1}{t+r/(1-r)}
\, - \, 1
\, - \, \frac{r}{(1-r)^2} \, \frac{1}{\left[ t + r/(1-r) \right]^2}
\right\},
\nonumber
\eea
where we have used the fact that
${\bf T_b}^2 \, = \, {\bf T_s}^2 \, = \, {\bf T_b \cdot T_s} \, = \, C_F = 4/3$
with $C_F$ the usual Casimir of the fundamental representation of $SU(3)$,
$E_g$ is the soft-gluon energy and $t$ is a unitary angular variable
\begin{equation}
t \, \equiv \, \frac{1 - \cos\vartheta}{2}
\end{equation}
with $\vartheta$ the emission angle of the gluon with respect to
the strange quark (in the soft limit there is no recoil and
the direction of the strange quark is not modified by gluon emission).
The mass-correction parameter is given in this case by
\begin{equation}
r \, \equiv \, \frac{m_s^2}{m_b^2}.
\end{equation}
Note that in the massless limit for the strange quark, the last term
on the r.h.s. of eq.~(\ref{square}) identically vanishes.
The ordinary quark velocity and the Lorentz factor are given
respectively by:
\beq
v \, \equiv \, \frac{p_s}{E_s} \, = \, \frac{1-r}{1+r};
~~~~~~~
\gamma \, \equiv \, \frac{E_s}{m_s} \, = \, \frac{1+r}{2 \sqrt{r}}.
\eeq
By taking the ultra-relativistic limit $r \ll 1$, we obtain:
\beq
- \, g_{\mu\nu} \, J^{\mu}(k) \, J^{\nu}(k)^*
\, \cong \, 4\pi \alpha \frac{C_F}{E_g^2} \left[ \frac{1}{t + r} \,
- \, 1 \, - \, \frac{r}{(t + r)^2} \right].
\eeq
We now introduce the approximations
\beq
\label{approxs}
\frac{1}{t+r} \, \simeq \, \frac{\theta(t - r)}{t},
~~~~~~
\frac{1}{(t+r)^2} \, \simeq \,
\theta(t - r) \, \frac{ 1 }{t^2},
\eeq
to extract the leading
behavior of the matrix element squared for $r \ll 1$
\footnote{
Eqs.~(\ref{approxs}) are intended to hold in an integral way:
\beq
\int_0^1 \frac{dt}{t+r} \, = \, \int_r^1 \frac{dt}{t} \,
+ \, O(r);
~~~~~
\int_0^1 dt \, \frac{1}{(t+r)^2} \, = \, \int_r^1
dt \frac{1}{t^2} \, + \, O(r).
\eeq
}.
We then obtain:
\beq
- g_{\mu\nu} J^{\mu}(k) J^{\nu}(k)^*
\simeq \,
4\pi \alpha \frac{C_F}{E_g^2}  \left[ \frac{\theta(t - r)}{t} - 1 -
 \frac{r \, \theta(t - r)}{t^2} \right]
 \simeq
4\pi \alpha \frac{C_F}{E_g^2} \theta(t - r) \left[ \frac{1}{t} - 1
 - \frac{r}{t^2} \right].
\eeq
On the last member we have
multiplied the term independent from $t$  (related to an isotropic
soft gluon emission off the beauty quark at rest) by the $\theta$
function, as this only introduces $O(r)$ terms and allows a
factorization of this ``dynamical'' constraint.

The approximations (\ref{approxs}) actually constitute the dead-cone
approximation, which is
the following limitation on lower emission angles of the gluon
with respect to the strange quark:
\begin{equation}
\vartheta \, > \, \vartheta_{\min} \, \equiv \, \frac{m_s}{E_s^{(0)}},
\end{equation}
where $E_s^{(0)}$ is the strange quark energy in lowest order, which can be identified
with the jet energy in the soft limit
\beq
E_{Xs} \simeq \, E_s^{(0)} \, = \, \frac{m_b}{2} \left( 1 \, + \, r \right),
\eeq
and $m_b$ is the beauty (pole) mass.
We then obtain for the angular variable the corresponding limitation
\begin{equation}
t \, > \, t_{\min} \, = \, \frac{1 - \cos\vartheta_{\min}}{2} \, \simeq \,
\left( \frac{\vartheta_{\min}}{2} \right)^2 \, \simeq \, r.
\end{equation}

\noindent
An analogous simplification can be made for the kinematical constraint of given
jet mass, i.e. of fixed (cfr eq.~(\ref{eqdefy}))
\beq
y \, \equiv \, \frac{m_{Xs}^2 - m_s^2}{m_b^2 - m_s^2},
\eeq
with $m_{Xs}^2  \equiv (p_s + p_g)^2$.
Let us note that $y$ is a unitary variable equal to zero in the Born kinematics
point $m_{Xs} \, = \, m_s$.
For a jet containing one soft gluon:
\beq
y
\, = \,
(1-r) \, \omega
\left[
t \, + \, \frac{r}{1-r}
\right]
\, \cong \,
\omega \, (t + r)
\, \simeq \,
\omega \, t
~~~~~~~~~~~~~~ {\rm for} ~ \, t \, > \, r,
\eeq
where
\beq
\omega \, \equiv \, \frac{2 E_g}{m_b(1 - r)}
\eeq
is the normalized gluon energy.

To sum up: the soft-limit mass distribution for a jet
initiated by a massive quark can be obtained from the
corresponding massless formula by simply adding the term
explicitly proportional to $m_s^2$ and imposing the dead cone
effect:
\begin{equation}
\label{softonly2} \left.
\frac{1}{\Gamma} \frac{d\Gamma^{(1)}}{dy} \right|_{\rm soft} \, =
\, \int_0^1 d\omega \int_r^1 d t \left[
        \frac{ A_1 \alpha}{\omega \, t}
\, + \, \frac{ D_1 \alpha}{\omega} \, + \, \frac{ D_1 \, r
\alpha}{\omega \, t^2} \right] \, \big[ \delta(y - \omega t) \, -
\, \delta(y) \big].
\end{equation}
The first-order coefficients read \cite{sghedoni}: 
\beq 
A_1 \, = \, \frac{C_F}{\pi};
~~~~~~~ D_1 \, = \, - \, \frac{C_F}{\pi}. 
\eeq 
The first term
proportional to $D_1$ on the r.h.s. of eq.~(\ref{softonly2}) is
related to soft emission off the initial beauty quark, in its rest
frame, while the second one is related to soft emission off the
fast-moving strange quark.

Eq.~(\ref{softonly2}) misses the contribution from hard collinear gluon emission.
We assume that the latter can be obtained from the massless one \cite{sghedoni} by simply imposing
the dead cone effect. The complete one-gluon distribution therefore reads:
\beq
\label{generale}
\frac{1}{\Gamma} \frac{d\Gamma^{(1)}}{dy}
\, = \,
\int_0^1 d\omega
\int_r^1 d t
\left[
         \frac{ A_1 \alpha}{\omega \, t}
\, + \,  \frac{ D_1 \alpha}{\omega} \, + \,  \frac{ D_1 \alpha \,
r}{\omega \, t^2} \, + \,  \frac{ B_1 \alpha}{t} \right] \, \big[
\, \delta(y - \omega t) \, - \, \delta(y) \, \big], \eeq where
\beq B_1 \, = \,  - \, \frac{3}{4}\frac{C_F}{\pi}. \eeq
A check of our assumption is provided by the comparison with the fixed
order Feynman diagram computation of the spectrum given in
sec.~\ref{check1}
\footnote{
A more rigorous and direct derivation can be obtained by using the
massive splitting functions (see for example \cite{Catani:2002hc}).
}.
The event fraction is given by:
\begin{equation}
\label{evfrac}
R(y) \, \equiv \,
\int_0^y \frac{1}{\Gamma} \frac{d\Gamma}{d y'} \, dy'
\, = \, 1 \, + \, R^{(1)}(y) \, + \, \cdots,
\eeq
where:
\beq
R^{(1)}(y) \, =
\, - \,
\int_0^1 d\omega
\int_r^1 d t
\left[
        \frac{ A_1 \alpha}{\omega \, t}
\, + \, \frac{ D_1 \alpha}{\omega}
\, + \, \frac{ D_1 \alpha \, r}{\omega \, t^2}
\, + \, \frac{ B_1 \alpha}{t}
\right] \,
\theta(\omega \, t - y).
\end{equation}
Note that the formula above has the correct end-point value: $R(1) = 1$.

\subsubsection{Leading Order}

Let us start by considering the leading-order term proportional to
the coefficient $A_1$ in eq.~(\ref{generale}). 
To include leading higher-order effects, 
we replace the tree-level coupling with the running coupling evaluated at the
gluon transverse momentum squared \cite{Amati:1980ch} (see later):  
\beq 
\alpha \, \to \,
\alpha(k_{\perp}^2), 
\eeq 
where the gluon transverse momentum
squared is defined as \cite{catanitrentadue}:
\begin{equation}
k_{\perp}^2 \, \equiv \, (1-r)^2 \, m_b^2 \omega^2 t \, \simeq \,
E_g^2 \, \vartheta^2 ~~~~ {\rm for}~\vartheta \, \ll \, 1.
\end{equation} By integrating the leading-order term on the r.h.s. of
eq.~(\ref{generale}) and expressing the result in terms of the
gluon transverse momentum, we obtain: \bea \label{recover}
\int_0^1 \frac{d\omega}{\omega} \int_r^1 \frac{d t}{t} A_1
\alpha(k_{\perp}^2 ) \, \delta(y - \omega \, t) &=& \frac{1}{y}
\int_{m_b^2 y^2}^{ m_b^2 y \, \min\left[1,\,y/r\right] } \frac{d
k_{\perp}^2 }{ k_{\perp}^2 } A_1 \alpha\left( k_{\perp}^2 \right)
\nonumber\\
&\simeq&
\frac{1}{y}
\int_{m_b^2 y^2}^{ m_b^2 y^2/(y+r) }
\frac{d k_{\perp}^2 }{ k_{\perp}^2 }
A_1 \alpha\left( k_{\perp}^2 \right),
\eea
where on the last member a smooth interpolation between the two
regions specified by the minimum has been considered
according to the approximation
\footnote{In general, step approximations are more convenient to obtain simple
analytical results while continuous functions are best suited for
numerical purposes.}
$\max[y, r] \, \approx \, y + r$.
Eq.~(\ref{recover}) reduces in the frozen coupling case to
$- A_1 \alpha/y \log(y + r)$.
As anticipated in the introduction,
we have therefore two different regions:
\begin{enumerate}
\item
{\bf high jet mass} --- compared to quark mass $m_s$:
\beq
y \, \gg \, r.
\eeq
Up to the logarithmic accuracy we are interested in,
one can extend this region up to $y \, > \, r$.
Eq.~(\ref{recover}) reduces to the massless case:
\begin{equation}
\label{recover2}
\frac{1}{y}
\int_{m_b^2 y^2}^{ m_b^2 y }
\frac{d k_{\perp}^2 }{ k_{\perp}^2 }
A_1 \alpha\left( k_{\perp}^2 \right).
\end{equation}
In this region, the jet mass is so large that,
at the logarithmic level, no effect
is left of $m_s \ne 0$.
To first order, eq.~(\ref{recover2}) becomes
$-A_1 \alpha/y \log y$;
\item
{\bf low jet mass:}
\beq
y \, \ll \, r.
\eeq
Analogously to the previous case, one can actually extend this region up to $y < r$.
Eq.~(\ref{recover}) specializes to:
\begin{equation}
\label{ancora}
\frac{1}{y}
\int_{m_b^2 y^2}^{ m_b^2 y^2/r }
\frac{d k_{\perp}^2 }{ k_{\perp}^2 }
A_1 \alpha\left( k_{\perp}^2 \right).
\end{equation}
It is worth noting that the effect of a non zero mass, $m_s \, \ne \, 0$,
is a restriction on the {\it upper} gluon transverse momenta.
To first order, eq.~(\ref{ancora}) becomes $- A_1\alpha/y \log r$.
In this low jet-mass region, in agreement with physical intuition,
the $s$ quark mass $m_s$ completely screens the collinear singularity
and produces a logarithm of the quark mass $m_s$ normalized to the relevant
hard scale $m_b$.
\end{enumerate}

\subsubsection{Subleading effects}

Let us discuss in this section the derivation of the sub-leading
effects in eq.~(\ref{generale}):
\begin{enumerate}
\item
the first term proportional to $D_1$, related to soft emission off
the initial heavy parton at rest, reads:
\beq
\int_0^1 \frac{d\omega}{\omega}
\int_r^1 d t  \, D_1 \, \alpha(k_{\perp}^2 )
\, \delta(y - \omega \, t)
\, \simeq \,
D_1 \, \alpha\left(m_b^2 \, y^2\right) \, \frac{1}{y}.
\eeq
This term is not modified with respect to the massless
case, again as expected on the basis of physical intuition;
\item
the second term proportional to $D_1$, related to soft emission
off the massive and fast-moving strange quark, reads:
\bea
\int_0^1 \frac{d\omega}{\omega}
\int_r^1 \frac{d t}{t^2} r \, D_1 \, \alpha(k_{\perp}^2 )
\, \delta(y - \omega \, t)
&\simeq&
\alpha \left( \frac{m_b^2 y^2}{y+r}\right) D_1 \left( \frac{1}{y} - \frac{1}{y+r} \right)
\nonumber\\
&\simeq&
\alpha \left(\frac{m_b^2 y^2}{r}\right) D_1 \frac{\theta(r-y)}{y}.
\eea
This term therefore vanishes in the high jet-mass region $y \, \gg \, r$,
where we recover the massless case (in which this term
was absent from the very beginning);
\item
the term proportional to $B_1$, related to hard collinear emission
off the strange quark, reads:
\beq
\int_0^1 d\omega
\int_r^1 \frac{d t}{t} B_1 \alpha(k_{\perp}^2 )
\, \delta(y - \omega \, t)
\, \simeq \,
B_1 \, \alpha\left(\frac{m_b^2 y^2}{y+r}\right) \, \frac{1}{y+r}
\, \simeq \,
 B_1 \, \alpha\left(m_b^2 y \right) \, \frac{\theta(y-r)}{y},
\eeq
where we have neglected small terms beyond the logarithmic accuracy.
This term is regulated by the non-vanishing strange mass and, as expected,
we re-obtain the massless case with the infrared cutoff $r$ on $y$.
\end{enumerate}
By collecting the various contributions computed in the previous sections,
we obtain for the $O(\alpha)$ distribution:
\beq
\label{tocoincide}
\frac{1}{\Gamma} \frac{d\Gamma^{(1)}}{dy} \, = \, -
\, A_1\alpha \frac{\log(y + r)}{y} \, + \, 2 D_1 \alpha  \frac{1}{
y } \, + \, (B_1 - D_1) \alpha  \frac{1}{ y + r }.
\eeq
Within logarithmic accuracy, we can make a sharp approximation to
obtain:
\begin{equation}
\label{final_particular}
\frac{1}{\Gamma}\frac{d\Gamma^{(1)}}{dy} \, = \, \Bigg\{
\begin{array}{ll}
\, - \, A_1\alpha \log r  \frac{1}{ y }
\, + \, 2 D_1 \alpha  \frac{1}{ y  } ,
&  ~~~~~ {\rm ~for~} y < r;
\\
\, - \, A_1\alpha \frac{\log y}{y}
\, + \, (B_1+D_1)\alpha  \frac{1}{ y } ,
& ~~~~~ {\rm ~for~} y > r.
\end{array}
\end{equation}
The main effects of virtual corrections are included, as usual,
by replacing the above functions with corresponding plus distributions.
The event fraction reads in the high-mass region:
\begin{equation}
R(y) \, = \, 1 \, - \, \int_y^1
\frac{1}{\Gamma} \, \frac{d\Gamma}{d y'} \, d y'
\, = \, 1 \, - \, \frac{1}{2} A_1 \alpha \log^2 y \, + \, (B_1 + D_1) \, \alpha \, \log y
~~~~~~~~~~ (y \, > \, r).
\end{equation}
By imposing that the event fraction is continuous across $y \, = \, r$,
\footnote{
That is equivalent to using the plus distributions.} 
\beq
R(y \to r^-) \, = \, R(y \to r^+) \, = \,
1 \, - \, \frac{1}{2} A_1 \alpha \log^2 r \, + \, (B_1 + D_1) \, \alpha \, \log r,
\eeq
we obtain the $\log r$ terms in the low-mass region:
\begin{equation}
R(y) \, = \, 1 \, - \, A_1 \, \alpha \, \log y \, \log r
\, + \, \frac{A_1}{2} \, \alpha \, \log^2 r
\, + \, 2 D_1 \alpha \log y \, + \, (B_1 - D_1) \alpha \log r
~~~~~~~~~~~~ (y \, < \, r).
\end{equation}

\subsection{Check with fixed-order computation}
\label{check1}

We compare in this section our eq.~(\ref{evfrac}) with the $O(\alpha)$ 
Feynman diagram computation of ref.~\cite{aliegreub}, 
where the decay spectrum has been computed
by retaining the non-vanishing strange-quark mass. The relevant
contribution is the one coming from the magnetic penguin operator
\beq 
O_7 \, = \, \frac{e}{16\pi^2} m_{b,\overline{MS}}(\mu_b) \,
\bar{s} \, \sigma_{\mu\nu}  b \, F^{\mu\nu}. 
\eeq
$m_{b,\overline{MS}}(\mu_b)$ is the $b$ mass in the
$\overline{MS}$ scheme and $\mu_b \, = \, O(m_b)$ is the
renormalization scale. By omitting non-logarithmic terms, the
fixed-order (fo) distribution in eq.~(30) of the first reference in
\cite{aliegreub} reads: 
\beq 
\left.\frac{1}{\Gamma}
\frac{d\Gamma^{(1)}}{dy} \right|_{\rm fo} \, = \, - \, A_1\alpha
\frac{\log(y + r)}{y} \, + \, 2 D_1 \alpha  \frac{1}{ y } \, + \,
(B_1 - D_1)\alpha  \frac{1}{ y + r }. 
\eeq 
The above formula
exactly coincides with the first-order expression in
eq.~(\ref{tocoincide}) derived in the previous section. This
comparison confirms the validity of the inclusion of the
coefficients of leading and subleading contributions as given in
eq.~(\ref{generale}).

\subsection{$N$-space}
\label{Nspace}

In order to resum the distribution to all orders in $\alpha$, one
has to transform it to moment $N$-space. The invariant mass
distribution in $N$-space is defined as: 
\beq 
\frac{ \, \Gamma_N^{(1)}}{\Gamma} \, = \, \int_0^1 dy \, (1-y)^{N-1} \, \frac{1}{\Gamma}
\frac{ d\Gamma^{(1)} }{dy}. 
\eeq
The transform to $N$ moment space indeed allows us to obtain an expression more
suitable for the resummation of the logarithmic terms. As
discussed in \cite{catani1,catani2} for the case of the jet mass event variable, 
the Mellin transform yields to a factorized expression of the phase space
\footnote{
The factorization of matrix elements follows instead from the dynamics of QCD.}. 
That gives rise to expressions for the jet mass distribution which are
accurate at the next-to-next-to-leading order
\cite{catani1,catani2}. The resulting expressions can be resummed in straightforward way.
We will follow here the same procedure in order to resum the mass logarithms
in eq.~(\ref{generale}). 
The exponent in the resummed expression reproduces
the usual structure, common to several resummed variables,
as the one shown in refs.~\cite{catanitrentadue,catani1,catani2}.
In practice, one exponentiates the one-gluon distribution in $N-$space
to account for multiple emission: 
\beq 
\frac{\, \Gamma_N}{\Gamma} \,
= \, 1 \, + \, \frac{ \, \, \Gamma_N^{(1)} }{\Gamma} \, + \,
\cdots \, \Rightarrow \, \exp \left[ \frac{ \, \, \Gamma_N^{(1)}
}{\Gamma} \right]. 
\eeq 
Gluon branching, i.e. secondary emission,
is taken into account by evaluating the running coupling at the
gluon transverse momentum squared \cite{Amati:1980ch}: 
\beq 
\alpha \, \to \, \alpha(k_{\perp}^2), 
\eeq
The ``effective'' one-gluon distribution therefore reads: \bea
\frac{ \, \, \Gamma_N^{(1)} }{\Gamma} &=& \int_0^1 d y \,
(1-y)^{N-1} \int_0^1 d\omega \int_r^1 d t \bigg[
        \alpha(k_{\perp}^2) \frac{ A_1 }{\omega \, t}
\, + \, \alpha(k_{\perp}^2) \frac{ D_1 }{\omega} \, + \,
\alpha(k_{\perp}^2) \frac{ D_1 \, r}{\omega \, t^2} \, + \,
\nonumber\\
&&~~~~~~~~~~~~~~~~~~~~~~~~~~~~~~~~~~~~~~~~~~~~~~~~ \, + \,
\alpha(k_{\perp}^2) \frac{ B_1}{t} \bigg]
 \big[ \, \delta(y - \omega t) \, - \, \delta(y) \, \big].
\eea The virtual contributions can be ``transferred'' to the
moment kernel as usual: \bea \label{quasifin} \frac{ \, \,
\Gamma_N^{(1)} }{\Gamma} &=& \int_0^1 d y \left[ (1-y)^{N-1} - 1
\right] \int_0^1 d\omega \int_r^1 d t \bigg[
        \alpha(k_{\perp}^2) \frac{ A_1 }{\omega \, t}
 +  \alpha(k_{\perp}^2) \frac{ D_1 }{\omega}
 +  \alpha(k_{\perp}^2) \frac{ D_1 \, r}{\omega \, t^2}
 +
\nonumber\\
&&~~~~~~~~~~~~~~~~~~~~~~~~~~~~~~~~~~~~~~~~~~~~~~~~~~~~~~~~~~~~~ \,
+ \, \alpha(k_{\perp}^2) \frac{ B_1}{t} \bigg] \, \delta(y -
\omega t). \eea In the following sections we perform the
integrations of the various terms above.

\subsection{Resummation}
\label{exponen}

By putting all the pieces together, we obtain the jet-mass distribution
for the heavy flavor decay in $N$-space:
\bea
\hskip -1 truecm
\frac{\Gamma_N}{\Gamma}
& \simeq &
\exp
\int_0^1 \frac{dy}{y} \left[ (1-y)^{N-1}-1 \right]
\Bigg\{
\int_{m_b^2 y^2}^{m_b^2 y \min[1,y/r]}
\frac{dk_{\perp}^2}{k_{\perp}^2} A_1 \, \alpha\left(k_{\perp}^2\right)
\, + \, \theta(y-r) B_1 \, \alpha\left(m_b^2 y\right) \, + \,
\nonumber\\
&&~~~~~~~~~~~~~~~~~~~~~~~~~~~~~~~~~~~~~
\, + \, D_1 \, \alpha\left(m_b^2 y^2\right)
\, + \, \theta(r-y) D_1 \, \alpha\left(m_b^2 y^2/r\right)
\Bigg\}.
\eea
A consistent next-to-leading logarithmic (NLL) resummation can be realized 
as in \cite{catanitrentadue}. 
The first-order coefficients $A_1$, $D_1$ and $B_1$ have been explicitly computed 
in the previous section.
The remaining NLL terms are related to two-loop effects.
In order to take them into account, in practice one has to include: 
\begin{enumerate}
\item
the two-loop correction ($\propto \beta_1$) 
in the QCD coupling $\alpha$ in the leading term $A_1 \, \alpha$; 
\item
the two-loop correction to the first-order term $A_1\, \alpha$ by means of the 
replacement
\beq
A_1 \alpha \, \to \, A_1 \alpha \, + \, A_2 \alpha^2 \, .
\eeq
It is pretty well established that the latter function is universal -- 
in the framework of effective theories, it is the well-known cusp 
anomalous dimension \cite{Korchemsky:1987wg}
\be
\Gamma_{\rm cusp}(\alpha) \, = 
\, \Gamma_{\rm cusp}^{(1)} \alpha \, + \, \Gamma_{\rm cusp}^{(2)} \alpha^2 \, + \, \cdots \, .
\ee
\end{enumerate}
What about resummation in higher orders, i.e. NNLL and beyond?
The coefficients $A_1$, $A_2$, $B_1$ and $D_1$ represent the lowest
order terms of the functions:
\footnote{
A compilation of the known coefficients $A_i$, $B_i$ and $D_i$
in our normalization, with references to the original papers, can be found in
\cite{Aglietti:2005mb}.}
\beq
A(\alpha) \, = \,  \sum_{n=1}^{\infty} A_n \, \alpha^n;
~~~~~
B(\alpha) \, = \,  \sum_{n=1}^{\infty} B_n \, \alpha^n;
~~~~~
D(\alpha) \, = \, \sum_{n=1}^{\infty} D_n \, \alpha^n.
\eeq
The functions $A(\alpha)$ and $B(\alpha)$ are related to small-angle emission only
and therefore represent universal intra-jet properties, as confirmed by explicit
higher-order computations \cite{eroi}.
The inclusion of the terms proportional to $\beta_2$, $A_3$ and $B_2$ is therefore
rather safe --- all these coefficients refer to universal ultraviolet or intra-jet quantities
\cite{Vogt:2000ci}. 

On the contrary, the function $D(\alpha)$, being related to soft emission at large angle, 
is in general a process-dependent inter-jet quantity.
In the framework of fragmentation functions, the function 
\be
D^{(f)}(\alpha) \, = \, D^{(f)}_1 \alpha \, + \, D^{(f)}_2 \alpha^2 \, + \, \cdots  
\ee
has been originally introduced in \cite{catcac},
where it was called $H(\alpha)$; the first-order coefficient $H_1$ was
also explicitly computed --- see the discussion in \cite{Cacciari:2002re}.
The $D$-function entering heavy flavor decays 
\be
D^{(h)}(\alpha) \, = \, D^{(h)}_1 \alpha \, + \, D^{(h)}_2 \alpha^2 \, + \, \cdots  
\ee
has however been shown to coincide with the former one to all orders \cite{gardi}:
\beq
D^{(f)}(\alpha) \, = \, D^{(h)}(\alpha) \, .
\eeq
The function $D^{(i)}(\alpha)$ with $i = f,h$ refers basically to soft radiation emitted by a 
heavy parton with a small virtuality.
In heavy flavor decays such as
\be
b \, \to \, X_s \, + \, \gamma \, ,
\ee
one considers a heavy flavor in its rest frame and looks
at the final invariant mass distribution, while in heavy flavor fragmentation
processes such as
\be
Z^0 \, \to \, b + \bar{b} \, ,
\ee
one has a fast-moving beauty quark and looks at its energy. 
Since the function $D(\alpha)$ is the same in both cases, that seems to imply
that the kinematical observable is irrelevant as long as soft-enhanced quantities
are concerned. The only thing that matters  is that of having a heavy quark close 
to its on-shell point point.
The first order coefficients of the above functions $D^{(f)}(\alpha)$ and $D^{(h)}(\alpha)$ 
coincide with our $D_1$.
As discussed in the introduction, our $D(\alpha)$ is related to massive partons but
it represents small-angle intra-jet corrections.
On the basis of physical intuition, we therefore conjecture that this coincidence extends 
to higher orders, i.e. that our $D(\alpha)$ coincides with $D^{(f)}(\alpha)$ or 
$D^{(h)}(\alpha)$.
From the above discussion it is clear that the inclusion of all the NNLL terms is not as 
rigorous as in the NLL case, the weakest point being the inclusion of the $D_2$ term.
An explicit check of our guess can be obtained by 
comparing our expanded resummation formula (given at the end of the paper) 
with an explicit (massive) two-loop computation, as soon as the latter becomes available.

Beyond NNLL approximation, only the coefficient $B_3$ is known analitically \cite{eroi}. 
We assume that the resummation formula keeps the same structure and that
by including (unknown) higher-order terms to our expansions,
\beq
A_1 \, \alpha \, + \, A_2 \, \alpha^2 \, + \, A_3 \, \alpha^3 \, \to \, A(\alpha),
~~~~~~~
B_1 \, \alpha \, + \, B_2 \, \alpha^2 \, \to \, B(\alpha),
~~~~~~~
D_1 \, \alpha \, + \, D_2 \, \alpha^2  \, \to \, D(\alpha),
\eeq
we obtain the decay spectrum formally resummed to all orders:
\bea
\hskip -1 truecm
\frac{\Gamma_N}{\Gamma}
& \simeq &
\exp
\int_0^1 \frac{dy}{y} \left[ (1-y)^{N-1}-1 \right]
\Bigg\{
\int_{m_b^2 y^2}^{m_b^2 y \min[1,y/r]}
\frac{dk_{\perp}^2}{k_{\perp}^2} A\left[ \alpha\left(k_{\perp}^2\right) \right]
 + \theta(y-r) B \left[\alpha\left(m_b^2 y\right)\right]
\nonumber\\
&&~~~~~~~~~~~~~~~~~~~~~~~~~~~~~~~~~~~~~~~
\, + \, D \left[\alpha\left(m_b^2 y^2\right)\right]
\, + \, \theta(r-y) \, D \left[\alpha\left(m_b^2 y^2/r\right)\right]
\Bigg\}.
\eea
In the limit $r \, \to \, 0$, we recover the well-known massless result.
It is remarkable that the single logarithmic terms $B$ and $D$
have $\theta$-functions of opposite arguments and therefore
exclude each other: either a soft contribution is present,
for a small jet-mass, or a collinear one is present, for a
high jet mass.
We now explicitly factor out the massless contribution in order to obtain
the mass-correction factor $\delta_N$ given in the introduction:
\bea
\hskip -1 truecm
\frac{\Gamma_N}{\Gamma}
&=&
\exp
\int_0^1 \frac{dy}{y} \left[ (1-y)^{N-1} - 1 \right]
\Bigg\{
\int_{ m_b^2y^2 }^{ m_b^2 y } \frac{dk_{\perp}^2}{k_{\perp}^2} A\left[\alpha\left(k_{\perp}^2\right)\right]
\, + \, B \left[\alpha\left(m_b^2 y\right)\right]
\, + \, D \left[\alpha\left(m_b^2 y^2\right)\right]
\nonumber\\
&+&
\theta(r-y) \,
\bigg[
- \, \int_{ m_b^2 y^2/r }^{ m_b^2 y } \frac{dk_{\perp}^2}{k_{\perp}^2} A\left[\alpha\left(k_{\perp}^2\right)\right]
- B \left[\alpha\left( m_b^2 y \right)\right]
+ D \left[\alpha\left( m_b^2 y^2/r \right)\right]
\bigg]
\Bigg\}.
\eea
In the massive case, one has therefore the additional factor:
\footnote{
A compact derivation of the correction factor $\delta_N$ is obtained by
splitting the angular integral in all the terms on the r.h.s. of eq.~(\ref{quasifin})
but the term proportional to $D_1 r$, as:
\beq
\int_r^1 dt = \int_0^1 dt - \int_0^r dt.
\eeq
The first integral on the r.h.s. represents the massless case while
the second one the mass corrections.
The unitary range in the latter is restored by setting $t'=t/r$.
}
\bea
\delta_N(m_b^2; \, m_s^2) &=&
\exp
\, \int_0^r \frac{dy}{y} \left[ (1-y)^{N-1} - 1 \right]
\Bigg\{
\, - \, \int_{ m_b^2 y^2/r }^{ m_b^2 y } \frac{dk_{\perp}^2}{k_{\perp}^2} A\left[\alpha\left(k_{\perp}^2\right)\right]
\, +
\nonumber\\
&&~~~~~~~~~~~~~~~~~~~~~~~~~~~~~~~
\, - \, B \left[\alpha\left( m_b^2 y \right)\right]
\, + \, D \left[\alpha\left( m_b^2 y^2/r \right)\right]
\Bigg\}.
\eea
The integral above can be transformed back to unitary range by means of
the rescaling $v \, = \, y/r,$
which gives:
\bea
\delta_N(m_b^2; \, m_s^2) &=&
\exp
\, \int_0^1 \frac{d v}{v} \left[ (1 -r \, v)^{N-1} - 1 \right]
\Bigg\{
\, - \, \int_{ m_s^2 v^2 }^{ m_s^2 v }
\frac{dk_{\perp}^2}{k_{\perp}^2} A\left[\alpha\left(k_{\perp}^2\right)\right]
\, +
\nonumber\\
&&~~~~~~~~~~~~~~~~~~~~~~~~~~~~~~~~~~~~~~~~
\, - \, B \left[\alpha\left( m_s^2 v \right)\right]
\, + \, D \left[\alpha\left( m_s^2 v^2 \right)\right]
\Bigg\}.
\eea
It is remarkable that, as a consequence of the rescaling, the hard scale
$Q$ (appearing in the limits of the transverse momentum integral
as well as in the argument of the coupling in the single-logarithmic terms)
changes from $m_b$ to $m_s$.
Let us also note that, in the last equation, the hard scale $m_b$ only enters through
the variable $r$ inside the moment kernel.

\noindent
To express the correction factor as a standard Mellin transform, we use the relation:
\beq
\label{rescale}
(1 - r \, y)^{N-1} \, - \, 1 \, = \, (1 - y)^{r(N-1)} \, - \, 1 \, + \, O \left( \frac{1}{N r} \right).
\eeq
Eq.~(\ref{rescale}) can be easily shown to be valid at the next-to-next-to-leading $\log N$ level,
by using the relation \cite{Aglietti:2002ew}
\beq
(1-y)^{N-1} \, - \, 1 \, \simeq \, - \, \theta\left(y-\frac{1}{n}\right)
\, + \, \frac{z(2)}{2} \,
\left[
\frac{1}{n} \, \delta \left(y - \frac{1}{n}\right)
\, - \, \frac{1}{n^2} \, \delta' \left(y - \frac{1}{n}\right)
\right]
\, + \, O\left[ N^3LO, \, \frac{1}{N} \right],
\eeq
where $n \equiv N e^{\gamma_E}$,
$\gamma_E \equiv \lim_{k \to \infty} \left[ \sum_{j=1}^k 1/j \, - \, \log k \right] \, = \, 0.577216 \cdots$
is the Euler constant, $z(a) \equiv \sum_{n=1}^{\infty} 1/n^a$ is the Riemann Zeta function
with $z(2) = \pi^2/6 = 1.64493\cdots$.
The final expression for the correction factor in $N$-space therefore reads:
\bea
\label{elei}
\delta_N(m_b^2; \, m_s^2) &=&
\exp
\, \int_0^1 \frac{d v}{v} \left[ (1 - v)^{r (N-1)} - 1 \right]
\Bigg\{
\, - \, \int_{ m_s^2 v^2 }^{ m_s^2 v }
\frac{dk_{\perp}^2}{k_{\perp}^2} A\left[\alpha\left(k_{\perp}^2\right)\right]
\, +
\nonumber\\
&&~~~~~~~~~~~~~~~~~~~~~~~~~~~~~~~~~~~~~~~~
\, - \, B \left[\alpha\left( m_s^2 v \right)\right]
\, + \, D \left[\alpha\left( m_s^2 v^2 \right)\right]
\Bigg\}.
\eea

\section{General case}
\label{generalsec}

This is the central section of the paper and contains general results
about threshold resummation in processes with jets initiated by partons
with a small mass compared to the hard scale.

\subsection{Mass Effects in a Jet}
\label{subsec1}

We generalize in this section the resummation formula obtained in
the previous section for the radiative $b$ decay by simply noting that any reference
to the particular process disappears in the correction factor $\delta_N$ in
eq.~(\ref{elei}).
Therefore we simply replace the beauty mass $m_b$ with the hard scale $Q$ of the general
process and $m_s$ with the mass $m$ of the quark triggering the jet under consideration:
\bea
\label{leiallafine}
\hskip -1truecm
\delta_N(Q^2; m^2) &=& \exp
\int_0^1 d z \, \frac{ z^{ \, r \, (N-1)} - 1 }{1-z}
\Bigg\{
 - \int_{ m^2 (1-z)^2 }^{ m^2 (1-z) } \frac{dk_{\perp}^2}{k_{\perp}^2} A\left[\alpha\left(k_{\perp}^2\right)\right]
 - B \left[\alpha\left( m^2 (1-z) \right)\right]
+ \nonumber\\
&&~~~~~~~~~~~~~~~~~~~~~~~~~~~~~~~~~~~~~~~~~~~~~~~~~~~~~~~~~~~~~~
\, + \, D \left[\alpha\left( m^2 (1-z)^2 \right)\right]
\Bigg\},
\eea
where $r$ is now defined in eq.~(\ref{defr}).
A check of our generalization is provided by the comparison with the full $O(\alpha)$ computation
of the DIS cross section with a massive quark in the final state (see next section).

As anticipated in the introduction, the jet mass distribution is an
infrared safe quantity, i.e. it has a finite limit for vanishing
quark mass, $m\to 0$.
Our results are in agreement with this general fact in the following way:
for $r \ne 0$, the distribution contains mass logarithms $\sim \log r$
which are not power suppressed, but have support in the region of power-suppressed
size $y < r$
\footnote{
Let us note that logarithmic mass effects only occur for $r N\gg 1$ because,
for $r N\ll 1$, one can expand the exponent in (\ref{leiallafine}) as
\beq
z^{ \, r \, (N-1)} \, - \, 1 \, = \, r(N-1) \log z \, + \, \frac{1}{2} r^2(N-1)^2 \log^2 z \, + \, \cdots,
\eeq
obtaining power-suppressed effects of the usual form.
The use of the resummation formula in the latter case however is not legitimate.
}.
The $O(\alpha)$ term in the correction factor in physical space $\delta(y; \, Q^2, m^2)$
can be obtained by subtracting out line 2 (the massless distribution)
from line 1 (the massive distribution) in eq.~(\ref{final_particular}):
\beq
\label{onlyfirst}
\delta(y;\,Q^2, m^2) \, = \,
\delta(y)
\, - \, A_1 \alpha \, \left[ \theta(r-y) \, \frac{\log r/y }{y} \right]_+
\, + \, (D_1 - B_1) \alpha \left[ \frac{\theta(r-y)}{y} \right]_+
\, + \, O(\alpha^2).
\eeq
Let us note that $\delta$ actually vanish for $y \, > \, r$
(where $r \ll 1$).

\subsection{Check with DIS with massive final quark}
\label{subsec2}

The first order corrections to the inclusive cross section have
been computed in \cite{Gott} for the charm production in
charged-current DIS, 
\beq 
\label{DISmassivo} 
\nu_{\mu} \, + \, s
\, \to \, \mu \, + \, c \, \, + \, (g), \eeq where $(g)$ is a real
or a virtual gluon. In this computation, the non zero charm mass
has been retained while the (much smaller) strange quark mass has
been neglected: \beq m_c \, = \, m \, \ne \, 0; ~~~~~ m_s \, = \,
0. \eeq Omitting non-logarithmic terms, the cross section given in
eq.~(40) of \cite{Gott} can be written as: 
\bea 
\label{diseq}
\frac{1}{\sigma} \frac{d\sigma}{dx} &\simeq& \delta(1-x) \, - \,
\frac{\alpha}{2\pi} \, \frac{1}{\bar{\epsilon}} \, P^{(0)}_{qq}(x)
\, + \, \frac{\alpha}{2\pi} \, P^{(0)}_{qq}(x) \, \log
\frac{Q^2}{\mu_F^2} \, +
\nonumber\\
&+& \frac{C_F \alpha}{\pi} \left[\frac{2\log(1-x) - \log(1 - \lambda \, x) }{1-x}\right]_+
- \frac{C_F \alpha}{\pi} \frac{1}{ \left[ 1 - x \right]_+ }
+ \frac{1}{4} \, \frac{C_F \alpha}{\pi} \frac{1}{\left[1 - \lambda \, x \right]_+},
\eea
where
\beq
\lambda \, \equiv \, \frac{Q^2}{Q^2 + m^2},
\eeq
$x = x_B \equiv Q^2/(2p\cdot q)$ is the Bjorken variable,
$Q^2 \equiv -q^2$ is the hard scale squared,
$q = p_{\mu} - p_{\nu_{\mu}}$ is the $W$-boson momentum, $\mu_F$ is the unit of mass of
dimensional regularization (to become the factorization scale after pole subtraction),
$P^{(0)}_{qq}(x)$ is the leading-order (massless) $q \to q$ splitting function
in 4 dimensions,
\beq
P^{(0)}_{qq}(x) \, \equiv \, C_F \left[ \frac{1+x^2}{1-x} \right]_+
\eeq
and
\beq
\frac{1}{\bar{\epsilon}} \, \equiv \, \frac{1}{\epsilon}
\, - \, \gamma_E \, + \, \log(4\pi),
\eeq
with $d \, = \, 4-2\epsilon$ the space-time dimension.
For simplicity's sake, let us set from now on $\mu_F \, = \, Q$.
The infrared pole, of collinear nature, is absorbed into
the quark non-singlet distribution function, which
reads, in the $\overline{MS}$ scheme:
\beq
F_{q/q}(x; \, Q^2) \, = \, \delta(1 - x) \, - \,
\frac{\alpha}{2\pi} \, \frac{1}{\bar{\epsilon}} \, P^{(0)}_{qq}(x) \, + \, O(\alpha^2)\,.
\eeq
The next step is to factorize from the cross section in eq.~(\ref{diseq})
the (massless) coefficient function:
\beq
C_{DIS}(x; \, Q^2) \, = \,
\delta(1-x) \, + \,A_1 \, \alpha \, \left[ \frac{\log (1-x)}{1-x} \right]_+
\, + \, B_1\, \alpha \, \frac{1}{ \left[ 1-x \right]_+ }
\, + \, O\left(\alpha^2\right).
\eeq
The latter is obtained as the following convolution
\cite{catanitrentadue,Catani:1990rr,catcac}
\beq
C_{DIS}(x; \, Q^2) \, = \, \Delta(x; \, Q^2) \, \otimes \, J(x; \, Q^2),
\eeq
where
\beq
\Delta(x; \, Q^2) \, = \, \delta(1-x) \, + \, 2 A_1 \, \alpha \, \left[ \frac{\log (1-x)}{1-x} \right]_+
\, + \, O\left(\alpha^2\right),
\eeq
is the radiative factor related to the observed initial-state jet, produced by the massless $s$
quark, while
\beq
J(x; \, Q^2) \, = \, \delta(1-x) \, - \,A_1 \, \alpha \, \left[ \frac{\log (1-x)}{1-x} \right]_+
\, + \, B_1\, \alpha \, \frac{1}{ ~ \left[ 1-x \right]_+ }
\, + \, O\left(\alpha^2\right)
\eeq
is the jet factor related to the unobserved final-state jet, initiated by the $c$
quark, in the massless approximation.
The cross section can therefore be written as:
\beq
\label{capito}
\frac{1}{\sigma} \frac{d\sigma}{dx}
\, \simeq \,
F(x ; \, Q^2) \otimes \Delta(x; \, Q^2) \otimes J(x; \, Q^2) \otimes
h(x; \, Q^2, m^2),
\eeq
where:
\bea
h(x; \, Q^2, m^2)
&=&
\delta(1-x) \, +
\, A_1 \, \alpha
\left\{
\left[ \frac{\log(1 - x)}{1-x} \right]_+
\, - \, \left[ \frac{\log(1 - x + r)}{1-x} \right]_+
\right\}
\, +
\nonumber\\
&& ~~~~~~~~~~~~~~~~~~~~~~
+ \, (D_1-B_1) \, \alpha
\left\{
\frac{1}{ \, \left[ 1-x \right]_+ }
\, - \, \frac{1}{\, \left[ 1 - x + r \right]_+ }
\right\}.
\eea
The ``$\otimes$'' denotes a convolution and we have used the fact that
\beq
\lambda \, = \, \frac{1}{1 + r}
\, \simeq \, 1 \, - \, r
\eeq
and
\beq
1 - \lambda \, x \, \cong \, 1 - x + r.
\eeq
The last term $h(x ; \, Q^2, m^2)$ on the r.h.s. of eq.~(\ref{capito}) is in agreement,
within logarithmic accuracy, with $\delta(y; \, Q^2, m^2)$
expanded to $O(\alpha)$, in eq.~(\ref{onlyfirst}), after setting $x = 1 - y$:
\beq
h(x; \, Q^2, m^2) \, \simeq \, \delta(1-x; \, Q^2, m^2).
\eeq
The resummation of mass effects in the charged-current DIS cross section, for several values
of the ratio $m_c/Q$, has been investigated in detail in \cite{Corcella:2003ib}.

\subsection{Tower Expansion}
\label{subsec3}

The universal mass-correction factor has the generalized exponential
structure \cite{catanitrentadue}
\beq
\delta_N(Q^2; \, m^2) \, = \, e^{ F_N \left( Q^2;m^2 \right) },
\eeq
where the exponent has a double expansion of the form:
\beq
F_N\left( Q^2; m^2 \right)
\, = \, \sum_{n=1}^{\infty} \sum_{k=1}^{n+1} F_{n k} \, \alpha^n \log^k (N r),
\eeq
with $F_{n k}$ numerical coefficients.
The exponent can be expanded in towers of logarithms as:
\bea
F_N\left( Q^2; m^2 \right)
&=&
\rho \, d_1 \left( \rho \right)
\, + \,
\sum_{n=0}^{\infty} \alpha^n \, d_{n+2}\left( \rho \right)
\nonumber\\
&=&
\rho \, d_1 \left( \rho \right)
\, + \, d_2 \left( \rho \right)
\, + \, \alpha \, d_3\left( \rho \right)
\, + \, \alpha^2 \, d_4\left( \rho \right)
\, + \, \cdots,
\eea
where
\beq
\rho \, \equiv \, \beta_0 \alpha(\mu^2) \, \log \left( N \, r \right)
\eeq
and $\mu = O(m)$ is a renormalization scale of the order of the quark mass $m$.

\noindent
By truncating the above series expansion, one obtains a fixed-logarithmic
approximation to the form factor $\delta_N$.
The functions $d_i(\rho)$, which represent the mass effects,
can be obtained from the standard ones $g_i(\lambda)$ \cite{Aglietti:2005mb}
by means of the replacements:
\footnote{
All these functions contain in principle the over-all factor
$\theta\left( N - 1/r \right)$, coming from the step approximation of the moment
kernel, which avoids modifications for small $N$ of the massless behaviour,
in agreement with physical intuition.
Analytic continuation to the complex $N$-plane is made by omitting such
factor. 
}
\beq
A(\alpha) \, \to \, - \, A(\alpha);
~~
B(\alpha) \, \to \, - \, B(\alpha);
~~
D(\alpha) \, \to \, D(\alpha);
~~
\log\frac{\mu^2}{Q^2} \, \to \, \log\frac{\mu^2}{m^2} ;
~~
\lambda \, \to \, \rho.
\eeq
It is worth observing that mass effects induce a similar
structure to the massless one, involving changes of sign of the collinear functions
$A$ and $B$, with the rescaling $Q \, \to \, m$.
We then obtain:
\begin{eqnarray}
d_1(\rho) &=&  \frac{{A_1}}{2\,\beta_0\, \rho}
\Big[
\left( 1 - 2\,\rho \right) \,
       \log (1 - 2\,\rho)
     - 2\,  \left( 1 - \rho \right)
           \,\log (1 - \rho)
\Big];
\\
d_2(\rho)&=& \frac{{D_1}}{2\,{{\beta}_0}} \log (1 - 2\,\rho)
- \frac{B_1}
   {{\beta}_0} \,\log (1 - \rho )
- \frac{{A_2}}{ 2\,
     {{\beta}_0}^2} \,
\Big[
\log (1 - 2\,\rho ) -
       2\,\log (1 - \rho )
\Big]  +
\\ \nonumber
&+&\frac{{A_1}\, {{\beta}_1} }{4\,{{{\beta}_0}}^3}\,
\Big[
       2\,\log (1 - 2\,\rho ) +
       {\log^2 (1 - 2\,\rho )} -
       4\,\log (1 - \rho ) -
       2\,{\log^2 (1 - \rho )}
\Big] \,
      +
\\ \nonumber
&-& \frac{{A_1}\,{{\gamma }_E}  }
     {{{\beta}_0}}\,
\Big[
\log (1 - 2\,\rho ) -
       \log (1 - \rho )
\Big]
- \frac{A_1}{2 \, \beta_0}
\Big[
\log(1 - 2\,\rho )-2 \,\log (1 -
\rho )
\Big]
\, \log \frac{\mu^2}{m^2}.
\end{eqnarray}
For the NNLO function $d_3$ we obtain:
\begin{eqnarray}
d_3(\rho) &=&
- \, \frac{D_2}{\beta_0} \,
\frac{\rho}{1 - 2\,\rho }
\, - \, D_1 \, \gamma_E \, \frac{2 \, \rho }{1 - 2\,\rho }
+ \frac{{D_1}\,{\beta_1}}{2\,
     {{\beta_0}}^2}\,
\left[
\frac{2\,\rho }{1 - 2\,\rho } +
       \frac{\log (1 - 2\,\rho )}
        {1 - 2\,\rho }
\right]
\, + \nonumber\\
&+& \frac{B_2}{\beta_0}
\, \frac{\rho }{ 1 - \rho }
\, + \, B_1 \, \gamma_E \, \frac{\rho }{1 - \rho }
\, + \nonumber\\
&-&
  \frac{ B_1 \, \beta_1 }{ {\beta_0}^2 } \,
\left[
\frac{\rho }{1 - \rho } +
       \frac{\log (1 - \rho )}{1 - \rho }
\right]
+ \frac{A_3}
     {2\,{{\beta_0}}^2}\,
\left[
\frac{\rho }{1 - 2\,\rho } -
       \frac{\rho }{1 - \rho }
\right]
+\nonumber\\
&+&  \frac{A_2\,
       {\gamma_E} }{
       {\beta_0}}\,
\left[
    \frac{2 \, \rho }{1 - 2\,\rho } -
         \frac{\rho}{1 - \rho }
\right]
+\nonumber\\
&-&
  \frac{ A_2 \, \beta_1 }{ 2\,{\beta_0}^3 }\,
\left[
\frac{3\,\rho }{1 - 2\,\rho } -
       \frac{3\,\rho }{1 - \rho } +
       \frac{\log (1 - 2\,\rho )}
        {1 - 2\,\rho } -
       \frac{2\,\log (1 - \rho )}{1 - \rho }
\right]
+\nonumber\\
&+&
  \frac{ A_1 \, {\gamma_E}^2 }{2}\,
\left[
\frac{4\,\rho }{1 - 2\,\rho } -
       \frac{\rho }{1 - \rho }
\right]
+ \frac{ A_1 \,{\pi}^2 }{12}\,
\left[
\frac{4\,\rho }{1 - 2\,\rho } -
       \frac{\rho }{1 - \rho }
\right]
+\nonumber\\
 &+&
  \frac{ A_1 \, \beta_2 }{4\,{\beta_0}^3}\,
\left[
\frac{2\,\rho }{1 - 2\,\rho } -
       \frac{2\,\rho }{1 - \rho } +
       2\,\log (1 - 2\,\rho ) -
       4\,\log (1 - \rho )
\right]
+\nonumber\\
&-& \frac{ A_1 \, \beta_1 \, \gamma_E }{ {\beta_0}^2 }\,
\left[
\frac{2\, \rho}{1 - 2\,\rho } -
       \frac{\rho}{1 - \rho } +
       \frac{\log (1 - 2\,\rho )}
        {1 - 2\,\rho } -
       \frac{\log (1 - \rho )}{1 - \rho }
\right]
+\nonumber\\
&+&
  \frac{A_1\,{\beta_1}^2}{ 2\, {\beta_0}^4 }\,
\left[
\frac{\rho }{1 - 2\,\rho } -
       \frac{\rho }{1 - \rho } -
       \log (1 - 2\,\rho ) +
       \frac{\log (1 - 2\,\rho )}
        {1 - 2\,\rho }
\right. \nonumber \\
&+&
\left.
       \frac{ \log^2 (1 - 2\,\rho ) }
        {2\,\left( 1 - 2\,\rho  \right) } +
       2\,\log (1 - \rho ) -
       \frac{2\,\log (1 - \rho )}
        {1 - \rho } -
       \frac{ \log^2 (1 - \rho ) }{1 - \rho }
\right]
\, + \, \nonumber\\
&-&
  \frac{D_1}{\beta_0}
\frac{\rho }{ 1 - 2\,\rho } \,
     \log \frac{{\mu }^2}{m^2}
+ \frac{ B_1}{\beta_0}
\frac{\rho }{ 1 - \rho }\,
     \log \frac{{\mu }^2}{m^2}
+ \frac{{A_2}}{{{\beta_0}}^2}\,
\left[ \frac{\rho }{1 - 2\,\rho } -
       \frac{\rho }{1 - \rho }
\right] \,
     \log \frac{{\mu }^2}{m^2}
+\nonumber\\
&+& \frac{{A_1}\,
     {\gamma_E} }{{\beta_0}}\,
\left[
\frac{2\,\rho }{1 - 2\,\rho } -
       \frac{\rho }{1 - \rho }
\right] \,
     \log \frac{{\mu }^2}{m^2}
+\nonumber\\
&-& \frac{{A_1}
     {\beta_1} }{{{\beta_0}}^3}
\left[
\frac{\rho }{1 - 2\,\rho } -
       \frac{\rho }{1 - \rho } +
       \frac{\log (1 - 2\,\rho )}{2} +
       \frac{{\rho }\,
          \log (1 - 2\,\rho )}{1 - 2\,\rho }
\right.
+ \, \nonumber\\
&-&
\left.
\log (1 - \rho )
- \frac{\rho \,\log (1 - \rho )}
        {1 - \rho }
\right]
     \log \frac{{\mu }^2}{m^2}\!
+\nonumber\\
&+& \frac{{A_1}}{2\,
     {\beta_0}}\,
\left[
\frac{2\,{\rho }^2}
        {1 - 2\,\rho } -
       \frac{{\rho }^2}{1 - \rho }
\right] \,
     {\log^2 \frac{{\mu }^2}{m^2}}.
\end{eqnarray}
The coefficients $\beta_i$ of the QCD $\beta$-function in our normalization
have been given in \cite{Aglietti:2005mb}.

\subsection{Inverse Mellin Transform}
\label{subsec4}

The mass-correction factor in physical space is obtained by means
of an inverse Mellin transform of $\delta_N$: 
\beq 
\delta\left(
y;\, Q^2, m^2 \right) \, = \, (1-y) \, \frac{d}{dy} \,
\left\{\int_{c-i\infty}^{c+i\infty} \frac{d N}{2\pi i N} \,
(1-y)^{-N} \, \delta_N\left( Q^2, m^2 \right) \right\}, 
\eeq 
where $c$ is a (real) constant chosen in such a way that the integration
contour lies to the right of all the singularities of $\delta_N$.
By defining 
\beq 
\bar{\delta}_{N r} \, \equiv \,\delta_{N} 
\eeq
and changing variable from $N$ to $\nu = N r$, we obtain: 
\beq
\delta\left( y; \, Q^2, m^2 \right) \, = \, \frac{d}{dy} \,
\left\{ \int_{c \, r - i\infty}^{c \, r + i\infty}
\frac{d\nu}{2\pi i\nu} \, \left[ ~ (1-y)^{1/r} \right]^{-\nu}
\bar{\delta}_{\nu} \left( Q^2, m^2 \right) \right\}, 
\eeq 
where we
have omitted the power correction $y$ multiplying the derivative
with respect to $y$. The correction factor in physical space is
therefore the inverse Mellin transform of $\bar{\delta}_\nu$ with
respect to $(1-y)^{1/r}$ 
\footnote{ 
Let us note that, in the large $N$ limit we are interested in (see eq.~(\ref{rescale})) 
$rN \gg 1$, we can make the approximation $r(N-1) \simeq r N-1$.
}. 
We can therefore use the results in \cite{Aglietti:2002ew} to
obtain the correction factor in physical space in NNLL
approximation: 
\beq 
\delta\left( y; \, Q^2, m^2 \right) \, = \,
\frac{d}{dy} \, \Big\{ \theta( y ) \, \Delta \left( y; \, Q^2, m^2
\right) \Big\} \eeq where:
\begin{equation}
\label{SigmaNNLO}
\Delta \left( y; \, Q^2, m^2 \right)
\,=\,
\frac{  e^{ L \, d_1(\tau) \, +  \, d_2(\tau)  }  }
{  \Gamma\left[1 - h_1(\tau) \right]  } \delta\Delta.
\end{equation}
We have defined
\begin{equation}
h_1(\tau) \, = \, \frac{d}{d\tau}\left[\tau d_1(\tau)\right]
\, = \, d_1(\tau) + \tau \, d_1'(\tau).
\end{equation}
$\delta \Delta$ is a NNLL correction factor which can be set
equal to one in NLL:
\begin{equation}
\delta\Delta_{NLL} \, = \, 1.
\end{equation}
Its NNLL expression reads:
\begin{equation}
\delta\Delta \,=\, \frac{ S ~ }{ ~~ S|_{L\rightarrow 0} }
\end{equation}
with
\begin{equation}
S \,=\, e^{ \alpha \, d_3(\tau) }
\Bigg\{
1 \, + \, \beta_0 \, \alpha \, d_2'(\tau) \,
\psi\left[ 1 - h_1(\tau) \right]
\, + \, \frac{1}{2} \beta_0 \, \alpha \, h_1'(\tau)
\Big\{
\psi^2\left[1-d_1(\tau)\right]
- \psi'\left[1-d_1(\tau)\right]
\Big\}
\Bigg\}.
\end{equation}
$\Gamma(x)$ is the Euler Gamma function and
\begin{equation}
\psi(x) \, \equiv \, \frac{d}{d x} \log \Gamma(x)
\end{equation}
is the digamma function.
Finally:
\beq
\tau \, \equiv \, \beta_0 \alpha \, L
\eeq
and
\beq
L \, \equiv \, - \, \log \left[ 1 - (1 - y)^{1/r} \right].
\eeq
It is convenient to approximate the argument of the
logarithm by an expansion for $y \ll r$:
\beq
(1-y)^{1/r} \, = \, 1 \, - \, \frac{y}{r} \, + \, O\left( \frac{y^2}{r^2}\right).
\eeq
Note that the r.h.s. is positive only for $y<r$, implying that the
linearization above shrinks the domain of $y$ from the unitary interval $(0,1)$
to the much smaller interval $(0,r)$.

To summarize, we have the final result:
\beq
\delta\left( y; \, Q^2, m^2
\right) \, = \, \frac{d}{dy} \, \Big\{ \theta( y ) \, \Delta\left( y; Q^2, m^2 \right)
\Big\}
\eeq
where $\Delta\left( y; \, Q^2, m^2 \right)$ is given in eq.~(\ref{SigmaNNLO}) and
\beq
L \, =
\, \theta(r-y) \, \log \frac{r}{y}.
\eeq
In agreement with the observation above, we have limited the domain to $y < r$
with a $\theta$-function.

\subsubsection{Coefficients of the mass logarithms}

The mass correction factor in physical space $\Delta\left( y; Q^2,m^2 \right)$
also has a generalized exponential structure:
\beq
\Delta\left( y; Q^2,m^2 \right)
 \, = \, e^{H\left( y; Q^2,m^2 \right)},
\eeq
where the exponent has a double expansion of the form:
\beq
H\left( y; Q^2, m^2 \right)
\, = \, \theta(r-y) \, \sum_{n=1}^{\infty}
\sum_{k=1}^{n+1} H_{n k} \, \alpha^n \log^k \frac{r}{y},
\eeq
with $H_{nk}$ numerical coefficients.
By expanding the r.h.s. of eq.~(\ref{SigmaNNLO}) up to third order, one obtains the
following expressions for the logarithmic coefficients:
\begin{eqnarray}
H_{12} &=& \frac{1}{2} A_1;
\\
H_{11} &=& B_1 - D_1;
\\
H_{23} &=& \frac{1}{2} A_1 \beta_0;
\\
H_{22} &=& \frac{1}{2} A_2 + \frac{1}{2} \beta_0 (B_1 - 2\,D_1) - \frac{1}{2} A_1^2 z(2);
\\
\label{refG21}
H_{21} &=& B_2 - D_2 - A_1 \left( B_1 - D_1 \right) z(2) - A_1^2 z(3);
\\
H_{34}&=& \frac{7}{12} A_1 \beta_0^2;
\\
H_{33}&=&  A_2 \beta_0 + \frac{1}{2}  A_1 \beta_1
+ \frac{1}{3}\beta_0^2 \left(  B_1 - 4\,  D_1 \right)
- \frac{3}{2}  A_1^2 \beta_0 z(2) -  \frac{1}{3}  A_1^3 z(3);
\\
\label{refG32}
H_{32}&=&
  \frac{1}{2} A_3
+ {\beta_0}\, ({B_2} - 2 \, D_2)
+ \frac{\beta_1}{2} \, (B_1 - 2 \, {D_1})
- {A_1} \, {A_2} \, z(2)
- \frac{ {A_1} \, {\beta_0} } {2}\, (5\,{B_1}\,
- 7 \, {D_1})\,z(2)
\, +
\nonumber\\
&-& \frac{ {{A_1}}^3 }{4} \, z(4)
- \frac{9 \, {{A_1}}^2 \, {\beta_0}\, {z}(3)}{2}
- {{A_1}}^2 \, ({B_1} \,- \, D_1) z(3),
\end{eqnarray}
where $z(3) = 1.20206\cdots$ and
$z(4) = \pi^4/90 = 1.08232\cdots$. Note that the leading
coefficients $H_{23}$ and $H_{34}$ involve products of the
one-loop coefficients $A_1$ and $\beta_0$ only.
The explicit expressions of the coefficients read:
\begin{eqnarray}
\label{Dij2}
H_{12} &=& \frac{C_F}{2\pi};
\\
H_{11} &=& \frac{C_F}{4\pi};
\\
H_{23} &=& \frac{C_F}{4 \pi^{2}} \left( \frac{11 C_A}{6} -\frac{n_f}{3}  \right);
\\
H_{22} &=& -\frac{C_F^2 z(2)}{2 \pi^2}+\frac{C_F}{4 \pi^{2}}\left[
C_A \left(\frac{433}{72}- z(2)\right) -\frac{35 n_f}{36}
 \right];
\\
H_{21} &=&\frac{ C_F^2}{2 \pi^2} \left(
+z(2)-\frac{3}{16}-5z(3)\right) \, +
\nonumber \\
&  & +\frac{C_F}{4 \pi^2}
\left[
n_f\left(\frac{239}{108}-\frac{2z(2)}{3}\right) + C_A \left(
\frac{-3595}{216}+\frac{5z(2)}{3}+19z(3)
\right)
\right];
\\
H_{34} &=& \frac{C_F}{48 \pi^3} \left[ C_A\left(\frac{847 C_A}{36}
-\frac{77 n_f}{9}\right)+\frac{7 n_f^2}{9} \right];
\\
H_{33} &=& -\frac{C_F^3 z(3)}{3 \pi^3}+\frac{C_F^2}{4
\pi^3}\left[-\frac{11 C_A z(2)}{2} +n_f(-\frac{1}{4}+z(2)) \right]
\nonumber \\
 & & + \frac{C_F}{4 \pi^3}\left[ \frac{11n_f^2}{36}+C_A^2 \left( \frac{1711}{144}-\frac{11z(2)}{6}
\right)+C_A n_f\left( -4+\frac{z(2)}{3}  \right) \right];
\\
 H_{32} & = & -\frac{C^3_F}{4 \pi^3} \left[z(4)+z(3)\right]+
\nonumber \\
 & &  +\frac{C^2_F}{4 \pi^3}
\left[\frac{-67n_f}{48}+\frac{61n_f z(2)}{36} +  5 n_f z(3)+ C_A
\left( -\frac{11}{32}-\frac{767 z(2)}{72}+2 z(2)^2-22 z(3) \right)
\right] +
\nonumber \\
& & + \frac{C_F}{4 \pi^3}\left[\frac{-3 n_f^2}{8} +\frac{n_f^2
z(2)}{9}+C_A^2 \left( \frac{-8855}{864}-\frac{145
z(2)}{36}+\frac{11 z(4)}{4}+\frac{319 z(3)}{12} \right) \right.+
\nonumber \\
& &  + C_A \left. \left(\frac{1549 n_f}{432}-\frac{35 n_f z(3)}{6}
\right)
\right],
\end{eqnarray}
where $C_A = N_c = 3$ and $n_f$ is the number of active quark flavors.

\section{Conclusions}
\label{conclusionsec}

We have resummed the invariant mass distribution of hadronic jets initiated
by massive quarks in next-to-leading logarithmic approximation.
The resummation has later been extended in heuristic way to the 
next-to-next-to-leading logarithmic approximation.
The mass effects have been relegated into a universal factor $\delta_N$ 
which takes into account the well-known dead-cone effect and soft radiation
characteristic of massive partons. $\delta_N$ contains the same
resummation functions which are encountered in standard threshold
resummation.
Mass corrections produce a universal intra-jet structure in agreement
with one's physical intuition: only small angle partons emissions are
involved, which can be ascribed to a specific jet in the event.
It is remarkable that the coefficients of the mass logarithms
are simply connected to those found in massless
processes, at low order as well as in higher order of perturbation
theory. A similar situation is found in the fragmentation of heavy quarks
\cite{catcac}.
Our formulae have been checked against explicit first-order
computations:
the radiative $b$ decay $b \to s \gamma$ with $m_s \neq 0$
and DIS $\nu_\mu + s \to c + \mu$ with $m_c \neq 0$,
finding complete agreement.

Mass effects, i.e. effects related to non vanishing parton masses,
often play a significant role in jet physics at the quantitative level
\cite{Krauss:2003cr}.
It may be worth to cite just a few applications of our results.

Semi-inclusive $B$-decays \beq \label{btoc} B \, \to \, X_c \, +
\, l \, + \, \nu_l \eeq are largely affected by the non-vanishing
charm quark mass as $r \gsim m_c^2/m_b^2 \approx 0.1$.
Semileptonic $b\to c$ decays allow the extraction of the CKM
matrix element $V_{cb}$ and constitute the main background to the
$b\to u l \nu$ decays, which are used for the extraction of the
CKM matrix element $V_{ub}$. The inclusion of charm mass effects
is needed to have a better understanding, for example, of the
charged lepton spectrum or the invariant hadron mass distribution,
which have recently been measured with great accuracy in
\cite{bcbabar,bc1,bc2,bc3,bc4,Urquijo:2006wd}. The resummed
formula which we have obtained can be combined with the full
$O(\alpha_S)$ triple differential distribution for (\ref{btoc})
recently obtained in \cite{Trott:2004xc,agru}. An additional
complication in this case stems from the fact that the charm quark
velocity, or equivalently the parameter $r$, is not fixed in the
tree-level process $b \to c l \nu$. That is because the hard
scale, set by the total hadron energy in the final state, is not
fixed. Since the exclusive channels $B \to D l \nu$ and $B \to D^*
l \nu$ constitute a large fraction of the total rate (\ref{btoc}),
an O(50\%) \cite{pdg}, the use of perturbation theory can be questioned.
We believe however that a perturbative computation can provide quantitative
informations on the decay (\ref{btoc}) to be compared with other
models.

An accurate computation of shape variable distributions in
$e^+e^-$ annihilations at the $Z^0$ pole and below --- such as
thrust, heavy jet mass, $C$-parameter, etc. --- asks for the
inclusion of the beauty mass effects very close to the two-jet
region. The mass correction parameter $r \approx 4 m_b^2/s \approx 0.1$
for $\sqrt{s} = 30$ GeV, while it is smaller by an order of magnitude
at the $Z^0$ pole.
At a future $e^+e^-$ linear collider of center-of-mass energy of 500 GeV, mass
effects in top pair production will be controlled by $r \approx
0.5$.

The invariant hadron mass distribution in semileptonic top decays,
\beq
t \, \to \, X_b \, + \, W
\eeq
is affected by the non-vanishing beauty mass close to threshold, i.e.
for $m_{Xb} \gsim m_b$.

In ref.~\cite{Cacciari:2005uk} a disagreement has been found in the (massless) evolution
of charm fragmentation data from $10$ to $91$ GeV.
We argue that the inclusion of charm mass effects, which should be significant for
$N \gsim (m_c/5)^{-2} \approx 10$, could improve the accuracy of the perturbative computations 
and eventually solve this problem.

To sum up, mass effects in a jet can be included, within logarithmic
accuracy,  by the universal factor in eq.~(\ref{maineq})
multiplying the massless jet function.

\vskip 0.6truecm

\centerline{\bf Acknowledgments}

We would like to thank F.~Ceccopieri for pointing out to us
ref.~\cite{Gott}.


\begin{thebibliography}{99}

\bibitem{qcdgeneral}
Y. Dokshitzer et al., {\it Basics of Perturbative QCD}, Editions
Frontieres, Paris (1991); 
R. Ellis, W. Stirling and B. Webber, {\it QCD and Collider Physics}, 
Cambridge University Press, Cambridge (1996). 

\bibitem{catanitrentadue}
S.~Catani and L.~Trentadue,
``Resummation Of The QCD Perturbative Series For Hard Processes,''
Nucl.\ Phys.\ B {\bf 327} (1989) 323.

\bibitem{Catani:1990rr}
S.~Catani, B.~R.~Webber and G.~Marchesini,
``QCD coherent branching and semiinclusive processes at large $x$,''
Nucl.\ Phys.\ B {\bf 349} (1991) 635.

\bibitem{catcac}
M.~Cacciari and S.~Catani,
``Soft-gluon resummation for the fragmentation of light and heavy quarks  at large $x$,''
Nucl.\ Phys.\ B {\bf 617} (2001) 253
[arXiv:hep-ph/0107138].

\bibitem{sghedoni}
U.~Aglietti, R.~Sghedoni and L.~Trentadue,
``Transverse momentum distributions in B decays,''
Phys.\ Lett.\ B {\bf 522} (2001) 83
[arXiv:hep-ph/0105322];
``Full $O(\alpha_s)$ evaluation for $b \to s \gamma$ transverse momentum
distribution,''
Phys.\ Lett.\ B {\bf 585} (2004) 131
[arXiv:hep-ph/0310360];
see also: R.~Sghedoni,
``Transverse momentum distributions in B decays,'' (Ph.~D.~thesis)
arXiv:hep-ph/0405291, last chapter.

\bibitem{Catani:2002hc}
S.~Catani, S.~Dittmaier, M.~H.~Seymour and Z.~Trocsanyi,
``The dipole formalism for next-to-leading order QCD calculations with massive partons,''
Nucl.\ Phys.\ B {\bf 627} (2002) 189
[arXiv:hep-ph/0201036].

\bibitem{Amati:1980ch}
D.~Amati, A.~Bassetto, M.~Ciafaloni, G.~Marchesini and G.~Veneziano,
``A Treatment Of Hard Processes Sensitive To The Infrared Structure Of QCD,''
Nucl.\ Phys.\  B {\bf 173} (1980) 429.

\bibitem{aliegreub}
A.~Ali and C.~Greub,
``Inclusive Photon Energy Spectrum In Rare $B$ Decays,''
Z.\ Phys.\ C {\bf 49} (1991) 431;
``Photon energy spectrum in $B \to X_s + \gamma$ and comparison with data,''
Phys.\ Lett.\ B {\bf 361} (1995) 146
[arXiv:hep-ph/9506374].

\bibitem{catani1}
S.~Catani, L.~Trentadue, G.~Turnock and B.~R.~Webber,
``Resummation of large logarithms in $e^+ e^-$ event shape distributions,''
Nucl.\ Phys.\  B {\bf 407} (1993) 3.

\bibitem{catani2}
S.~Catani, G.~Turnock, B.~R.~Webber and L.~Trentadue,
``Thrust Distribution in $e^+ e^-$ Annihilation,''
Phys.\ Lett.\  B {\bf 263} (1991) 491.

\bibitem{Korchemsky:1987wg}
G.~P.~Korchemsky and A.~V.~Radyushkin,
``Renormalization of the Wilson Loops Beyond the Leading Order,''
Nucl.\ Phys.\  B {\bf 283} (1987) 342.

\bibitem{Aglietti:2005mb}
U.~Aglietti, G.~Ricciardi and G.~Ferrera,
``Threshold resummed spectra in $B \rightarrow X_u\, l\, \nu$ decays in NLO\, (I)''
Phys.\ Rev.\ D {\bf 74} (2006) 034004
[arXiv:hep-ph/0507285].

\bibitem{eroi}
S.~Moch and A.~Vogt,
``Higher-order soft corrections to lepton pair and Higgs boson  production,''
Phys.\ Lett.\  B {\bf 631} (2005) 48 [arXiv:hep-ph/0508265].

\bibitem{Vogt:2000ci}
A.~Vogt,
``Next-to-next-to-leading logarithmic threshold resummation for
deep-inelastic scattering and the Drell-Yan process,''
Phys.\ Lett.\  B {\bf 497} (2001) 228 [arXiv:hep-ph/0010146].

\bibitem{Cacciari:2002re}
M.~Cacciari, G.~Corcella and A.~D.~Mitov,
``Soft-gluon resummation for bottom fragmentation in top quark decay''
JHEP {\bf 0212} (2002) 015 [arXiv:hep-ph/0209204].

\bibitem{gardi}
E.~Gardi, "On the quark distribution in an on-shell heavy quark and its all-order relations
with the perturbative fragmentation function", JHEP {\bf 0502} (2005) 53 [arXiv:hep-ph/05010257].

\bibitem{Aglietti:2002ew}
U.~Aglietti and G.~Ricciardi,
``NNLO threshold resummation in heavy flavour decays,''
Phys.\ Rev.\ D {\bf 66} (2002) 074003
[arXiv:hep-ph/0204125].

\bibitem{Gott}
T.~Gottschalk,
``Chromodynamic Corrections To Neutrino Production Of Heavy Quarks,''
Phys.\ Rev.\ D {\bf 23} (1981) 56.

\bibitem{Corcella:2003ib}
G.~Corcella and A.~D.~Mitov,
``Soft-gluon resummation for heavy quark production in charged-current deep
inelastic scattering,'' Nucl.\ Phys.\ B {\bf 676} (2004) 346
[arXiv:hep-ph/0308105];
for a general introduction, see also:
A.~Mitov, ``Applications of perturbative quantum chromodynamics to processes with
heavy quarks,'' (Ph.~D. thesis) arXiv:hep-ph/0311101.

\bibitem{Krauss:2003cr}
F.~Krauss and G.~Rodrigo,
``Resummed jet rates for $e^+e^-$ annihilation into massive quarks,''
Phys.\ Lett.\ B {\bf 576} (2003) 135
[arXiv:hep-ph/0303038].

\bibitem{bcbabar}
B.~Aubert {\it et al.}  [BABAR Collaboration],
``Measurement of the electron energy spectrum and its moments in inclusive $B\to X e \nu$
decays'', Phys.\ Rev.\ D {\bf 69} (2004) 111104
[arXiv:hep-ex/0403030].

\bibitem{bc1}
A.~H.~Mahmood {\it et al.}  [CLEO Collaboration],
``Measurement of the $B$-meson inclusive semileptonic branching fraction and
electron energy moments,''
Phys.\ Rev.\ D {\bf 70} (2004) 032003
[arXiv:hep-ex/0403053].

\bibitem{bc2}
K.~Abe {\it et al.}  [BELLE Collaboration],
``Moments of the electron energy spectrum in semileptonic $B$ decays at Belle,''
arXiv:hep-ex/0409015.

\bibitem{bc3}
K.~Abe {\it et al.}  [BELLE Collaboration],
``Moments of the hadronic mass spectrum in inclusive semileptonic $B$ decays at Belle,''
arXiv:hep-ex/0408139.

\bibitem{bc4}
D.~Acosta {\it et al.}  [CDF Collaboration],
``Measurement of the moments of the hadronic invariant mass distribution in
semileptonic $B$ decays,''
Phys.\ Rev.\ D {\bf 71} (2005) 051103
[arXiv:hep-ex/0502003].

\bibitem{Urquijo:2006wd}
P.~Urquijo, ``Moments of the electron energy spectrum and partial
branching fraction of $B \rightarrow X_c e \nu$ decays at Belle,''
arXiv:hep-ex/0610012.

\bibitem{Trott:2004xc}
M.~Trott, ``Improving extractions of $|V_{cb}|$ and $m_b$ from the hadronic invariant  
mass moments of semileptonic inclusive $B$ decay,''
Phys.\ Rev.\ D {\bf 70} (2004) 073003
[arXiv:hep-ph/0402120].

\bibitem{agru}
V.~Aquila, P.~Gambino, G.~Ridolfi and N.~Uraltsev,
``Perturbative corrections to semileptonic $b$ decay distributions,''
Nucl.\ Phys.\ B {\bf 719} (2005) 77
[arXiv:hep-ph/0503083].

\bibitem{pdg}
W.-M. Yao {\it et al.}, The Rewiew of Particle Physics, Journal of Physics G 33, 1 (2006). 

\bibitem{Cacciari:2005uk}
M.~Cacciari, P.~Nason and C.~Oleari,
``A study of heavy flavoured meson fragmentation functions in $e^+e^-$ annihilation,''
JHEP {\bf 0604}, 006 (2006)
[arXiv:hep-ph/0510032].

\end{thebibliography}
\end{document}